\documentclass[a4paper,11pt]{article}

\usepackage{jheppub}

\usepackage{amsfonts,graphics,epsfig,subfigure}

\title{An alternative perspective to observe the critical phenomena of dilaton AdS black holes}

\author[]{Jie-Xiong Mo}

 \affiliation[]{Institute of Theoretical Physics, Lingnan Normal University, Zhanjiang, 524048, Guangdong, China}

\emailAdd{mojiexiong@gmail.com}

\abstract{The critical phenomena of dilaton AdS black holes are probed from a totally different perspective other than the $P-v$ criticality and the $q-U$ criticality discussed in the former literature. We investigate not only
the two point correlation function but also the entanglement entropy of dilaton AdS black holes. We achieve this goal by solving the equation of motion constrained by the boundary condition numerically and we concentrate on $\delta L$ and $\delta S$ which have been regularized by subtracting the terms in pure AdS with the same boundary region. For both the two point correlation function and the entanglement entropy, we consider $4\times2\times2=16$ cases due to different choices of parameters. The van der Waals like behavior can be clearly witnessed from all the $T-\delta L$ ($T-\delta S$) graphs for $q<q_c$. Moreover, the effects of dilaton gravity and the spacetime dimensionality on the phase structure of dilaton AdS black holes are disclosed. Furthermore, we discuss the stability of dilaton AdS black holes by applying the analogous specific heat definition and remove the unstable branch by introducing a bar $T=T_*$. It is shown that the first order phase transition temperature $T_*$ is affected by both $\alpha$ and $n$. The analogous equal area laws for both the $T-\delta L$ graph and the $T-\delta S$ graph are examined numerically. The relative errors for all the cases are small enough that we can safely conclude that the analogous equal area laws hold for $T-\delta L$ ($T-\delta S$) graph of dilaton AdS black holes.}

\begin{document}
\maketitle
\flushbottom

\section{Introduction}
\label{sec:1}
 According to the well-known AdS/CFT correspondence \cite{ads1,ads2,ads3}, the physics in the bulk is related with the conformal field theory on the boundary one dimension fewer. Recently, Johnson disclosed intriguing properties of entanglement entropy \cite{Johnson} and further enhanced this relation. It was shown that the isocharges in the entanglement entropy-temperature plane and that of entropy-temperature plane not only enjoy similar behavior, but also share the same critical temperature and critical exponents \cite{Johnson}. Moreover, Ref. \cite{Nguyen} proved that equal area law, which was shown to hold for $T-S$ curve in former literature\cite{Spallucci}, also holds for the entanglement entropy-temperature plane. Ref. \cite{zengxiaoxiong3} further showed that other nonlocal observable, such as two point correlation function, also display similar behavior as the entanglement entropy. The research topic initiated by Johnson  is receiving more and more attention \cite{Caceres}-\cite{zhaoliu}.

  In this paper, we would like to generalize the above topic to dilaton AdS black holes to probe whether these phenomena are universal. On the other hand, our probe may be served as an alternative perspective to observe the critical phenomena of dilaton AdS black holes and would help deepen the understanding of these phenomena. Investigating the properties of AdS black holes in dilaton gravity is of interest itself. Dilaton field appears in the low energy limit of string theory and have significant impact on both the casual structure and thermodynamics of black holes. The action of dilaton gravity contains one or more Liouville-type potentials. These potentials are resulted by the breaking of spacetime supersymmetry in ten dimensions. Both the black hole solutions in dilaton gravity and their thermodynamics have attracted considerable attention \cite{Gibbons1}- \cite{xiong9}.

    The organization of this paper is as follows. Sec.~\ref{sec:2} devotes to a brief review of thermodynamics of dilaton AdS black holes.  In Sec.~\ref {sec:3}, we will study their two point correlation functions numerically while we will carry out numerical check of equal area law in Sec.~\ref{sec:4}. Furthermore, we will investigate their entanglement entropy numerically in Sec.~\ref{sec:5} and check the corresponding equal area law in Sec.~\ref{sec:6}. The last section devotes to conclusions.

\section{A brief review of thermodynamics of dilaton AdS black holes}
\label{sec:2}
 The $(n+1)$-dimensional Einstein-Maxwell-Dilaton action reads ~\cite{liweijia1,liweijia2,zhaoren}
\begin{equation}
S=\frac{1}{16\pi}\int d^{n+1}x \sqrt{-g}\left(R-\frac{4}{n-1}(\nabla \Phi)^2-V(\Phi)-(e^{-4\alpha \Phi/(n-1)}F_{\mu\nu}F^{\mu\nu}\right),\label{1}\\
\end{equation}
where $\Phi$ is the dilaton field with its potential denoted as $V(\Phi)$. $R$ is the Ricci scalar curvature and $F_{\mu\nu}=\partial_{\mu}A_{\nu}-\partial_{\nu}A_{\mu}$ is the electromagnetic field tensor. The strength of coupling between the electromagnetic field and the scalar field is characterized by $\alpha$.

The corresponding solution has been derived as~\cite{liweijia1,liweijia2,zhaoren}
\begin{eqnarray}
ds^2&=&-f(r)dt^2+\frac{1}{f(r)}dr^2+r^2R^2(r)d\Omega^2_{k,n-1},\label{2}\\
\Phi(r)&=&\frac{(n-1)\alpha}{2(\alpha^2+1)}\ln \left(\frac{b}{r}\right),\label{3}
\end{eqnarray}
where
\begin{eqnarray}
f(r)&=&-\frac{k(n-2)(1+\alpha^2)^2b^{-2\gamma}r^{2\gamma}}{(\alpha^2-1)(\alpha^2+n-2)}-\frac{m}{r^{(n-1)(1-\gamma)-1}}
\nonumber
\\
&\,&-\frac{n b^{2\gamma}(1+\alpha^2)^2r^{2(1-\gamma)}}{l^2(\alpha^2-n)}+\frac{2q^2(1+\alpha^2)^2b^{-2(n-2)\gamma}r^{2(n-2)(\gamma-1)}}{(n-1)(n+\alpha^2-2)},\label{4}\\
R(r)&=&e^{2\alpha \Phi(r)/(n-1)}.\label{5}
\end{eqnarray}
$k$ is a constant characterizing the hypersurface $d\Omega^2_{k,n-1}$ whose volume is denoted as $\omega_{n-1}$. $k$ can be taken as $-1,\,0,\,1$, corresponding to hyperbolic, flat and spherical constant curvature hypersurface respectively. $\gamma$ is related to $\alpha$ by $\gamma=\alpha^2/(\alpha^2+1)$. $b$ is an arbitrary constant while $l$ is the AdS length scale. $q$ and $m$ are constants related to the charge $Q$ and the mass $M$ of the black hole as follows
\begin{eqnarray}
 M&=&\frac{b^{(n-1)\gamma}(n-1)\omega_{n-1}m}{16\pi (\alpha^2+1)} ,\label{6}
\\
Q&=&\frac{\omega_{n-1}q}{4\pi}.\label{7}
\end{eqnarray}

Based on both the fact that the term including
 $m$ should vanish at spacial infinity and the fact that the electric potential $A_t$ should be finite at infinity, Ref.~\cite{Sheykhi8} obtained the restrictions on $\alpha$ for dilaton black holes coupled with power-law Maxwell field as follow
 \begin{eqnarray}
&\,&For \;\frac{1}{2}<p<\frac{n}{2},\;\;\;0\leq\alpha^2<n-2,\nonumber
\\
&\,&For \;\frac{n}{2}<p<n-1,\;\;\;2p-n<\alpha^2<n-2.\label{8}
\end{eqnarray}
Here, we consider the dilaton black holes coupled with standard Maxwell field in this paper, corresponding to the case $p=1$. So the above restriction reduces to be $0\leq\alpha^2<n-2$.

The Hawking temperature and the entropy have been derived as~\cite{zhaoren}
\begin{eqnarray}
T&=&\frac{-(1+\alpha^2)}{2\pi(n-1)}\Big[\frac{k(n-2)(n-1)}{2b^{2\gamma}(\alpha^2-1)}r_+^{2\gamma-1}+\Lambda b^{2\gamma} r_+^{1-2\gamma}+q^{2}b^{-2(n-2)\gamma}r_+^{(2n-3)(\gamma-1)-\gamma}\Big],\label{9}\\
S&=&\frac{\omega_{n-1}b^{(n-1)\gamma}r_+^{(n-1)(1-\gamma)}}{4},\label{10}
\end{eqnarray}
where the relation $\Lambda=-n(n-1)/2l^2$ holds for $(n+1)$-dimensional AdS black holes.

Furthermore, Ref.~\cite{zhaoren} investigated the $P-v$ criticality of dilaton AdS black holes when $Q$ was treated as an invariant parameter and the $q-U$ criticality ($U$ denotes the electric potential) when $l$ was treated as an invariant parameter. Both cases were shown to exhibit the van der Waals like behavior~\cite{zhaoren}. In the rest of this paper, we will further probe the critical phenomena of dilaton AdS black holes from an alternative perspective. Namely, the two point correlation function and the entanglement entropy.

\section{Two point correlation function of dilaton AdS black holes and its van der Waals like behavior}
\label{sec:3}
Considering the points $(t_0,
x_i)$ and $(t_0, x_j)$ on the AdS boundary with the corresponding bulk geodesic length denoted as $L$, the equal time two point correlation function in the large $\Delta$ limit ($\Delta$ denotes the conformal dimension of scalar operator $\cal{O}$ in the dual field theory) takes the following form \cite{Balasubramanian61}
 \begin{equation}
\langle {\cal{O}} (t_0,x_i) {\cal{O}}(t_0, x_j)\rangle  \approx
e^{-\Delta {L}}.\label{ll}
\end{equation}

The proper length can be obtained by parameterizing the trajectory with $\theta$
\begin{eqnarray}
L=\int_0 ^{\theta_0}\mathcal{L}(r(\theta),\theta) d\theta,~~\mathcal{L}=\sqrt{\frac{r'^2}{f(r)}+r^2}, \label{12}
 \end{eqnarray}
where $r'=dr/ d\theta$. Note that the boundary points have been chosen as $(\phi=\frac{\pi}{2},\theta=0)$ and $(\phi=\frac{\pi}{2},\theta=\theta_0)$.

Applying the well-known Euler-Lagrange equation
\begin{equation}
\frac{\partial \mathcal{L}}{\partial r}=\frac{d}{d\theta}\left(\frac{\partial \mathcal{L}}{\partial r'}\right), \label{13}
\end{equation}
the equation of motion for $r(\theta)$ can be derived. Solving the equation of motion constrained by the boundary condition $r(0)= r_0, r'(0)=0$, one can obtain $r(\theta)$ via numerical methods. To avoid the divergence, the geodesic length $L_0$ in pure AdS  with the same boundary region should be subtracted from the geodesic length $L$, with the regularized  geodesic length denoted as $\delta L$. Note that $L_0$ can also be obtained through numerical treatment.

To investigate the effect of dilaton gravity on the phase structure of two point correlation function, we focus on the cases that $\alpha$ is chosen as $0, 0.25, 0.5, 0.75$ respectively. Note that we have considered the restriction $0\leq\alpha^2<n-2$ mentioned in the former section. The case $\alpha=0$ is chosen so that one can compare the results with those of RN-AdS black holes.

On the other hand, $n$ is chosen as $3,4$ respectively to study the possible effect of spacetime dimensionality. Moreover, we choose $\theta_0=0.2, 0.3$ to check whether different boundary region sizes exert influence on the phase structure just as shown in former literature \cite{zengxiaoxiong3}. In this paper the cutoff $\theta_c$ will be chosen as $0.199, 0.299$ accordingly and the AdS radius $l$ will be set to be one.

\begin{figure*}
\centerline{\subfigure[]{\label{1a}
\includegraphics[width=8cm,height=6cm]{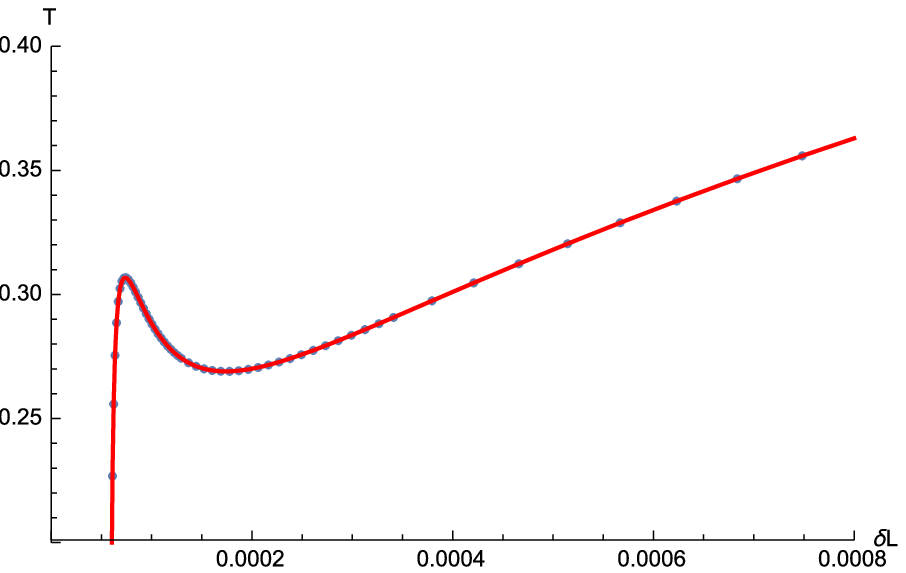}}
\subfigure[]{\label{1b}
\includegraphics[width=8cm,height=6cm]{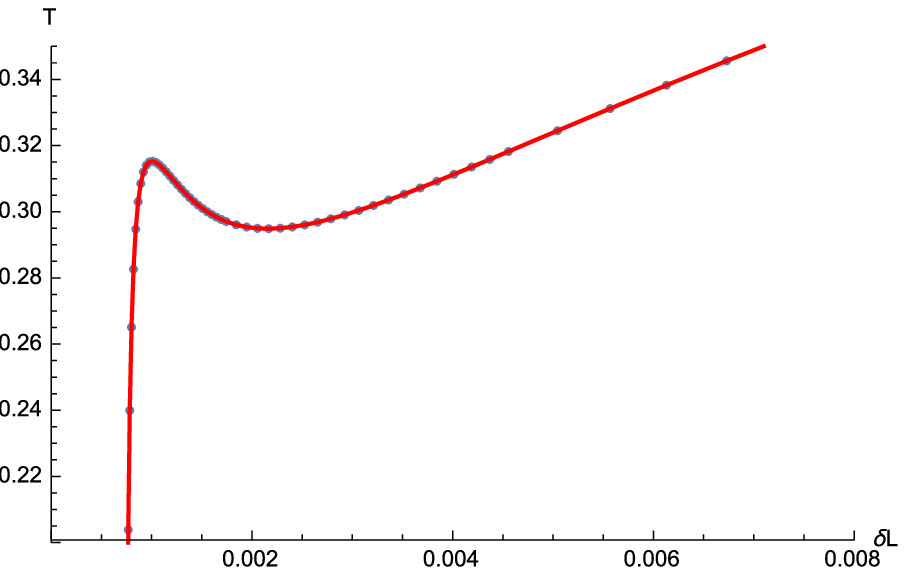}}}
\centerline{\subfigure[]{\label{1c}
\includegraphics[width=8cm,height=6cm]{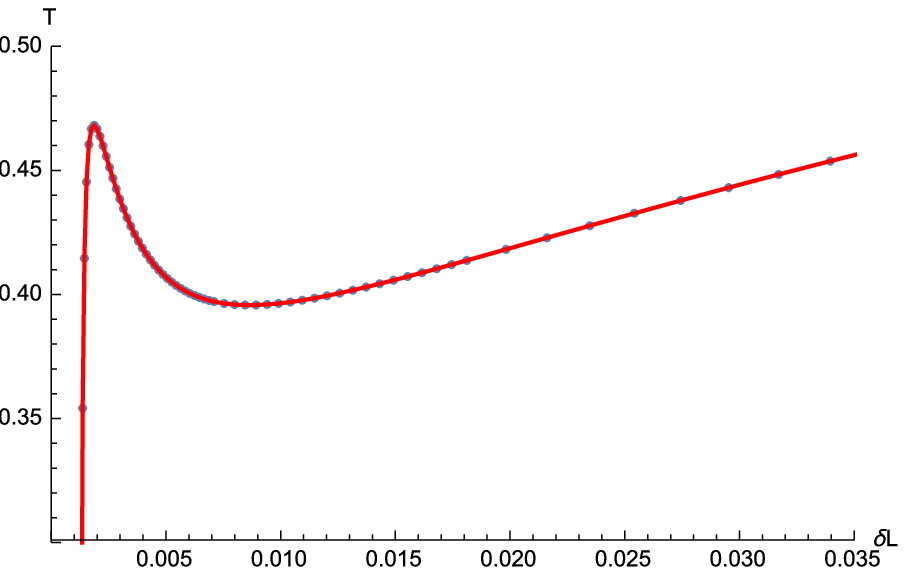}}
\subfigure[]{\label{1d}
\includegraphics[width=8cm,height=6cm]{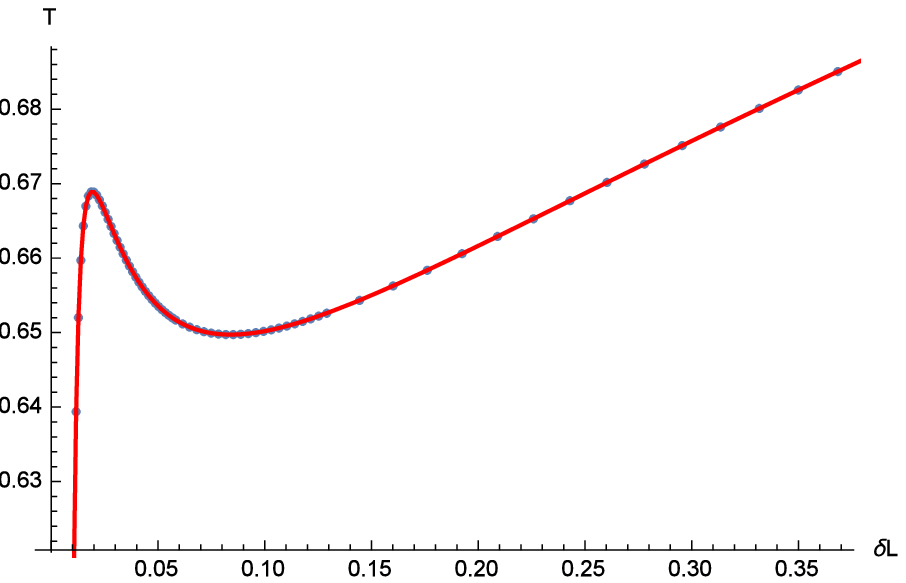}}}
 \caption{$T$ vs. $\delta L$ for $n=3, \theta_0=0.2$ (a) $\alpha=0, q=0.12$ (b) $\alpha=0.25, q=0.12$ (c) $\alpha=0.5, q=0.06$ (d) $\alpha=0.75, q=0.06$} \label{fg1}
\end{figure*}

\begin{figure*}
\centerline{\subfigure[]{\label{2a}
\includegraphics[width=8cm,height=6cm]{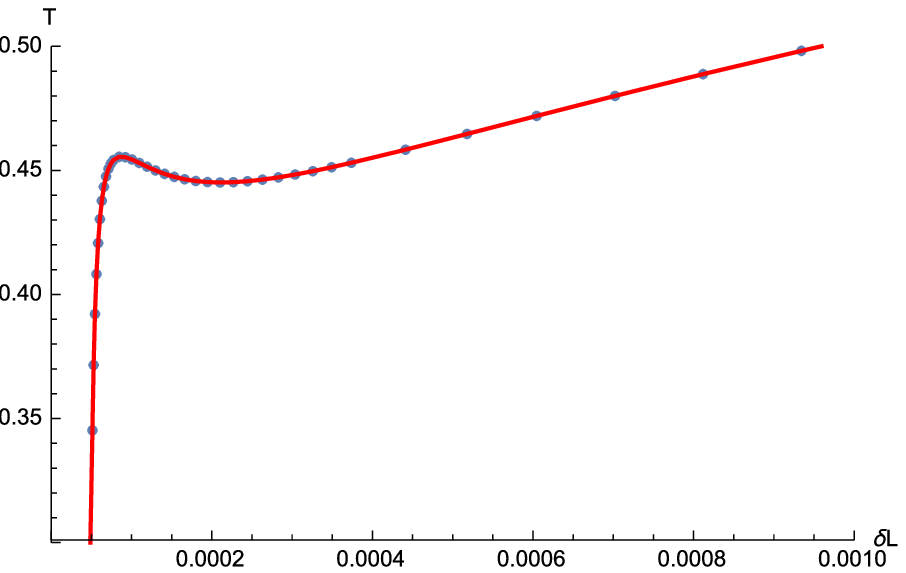}}
\subfigure[]{\label{2b}
\includegraphics[width=8cm,height=6cm]{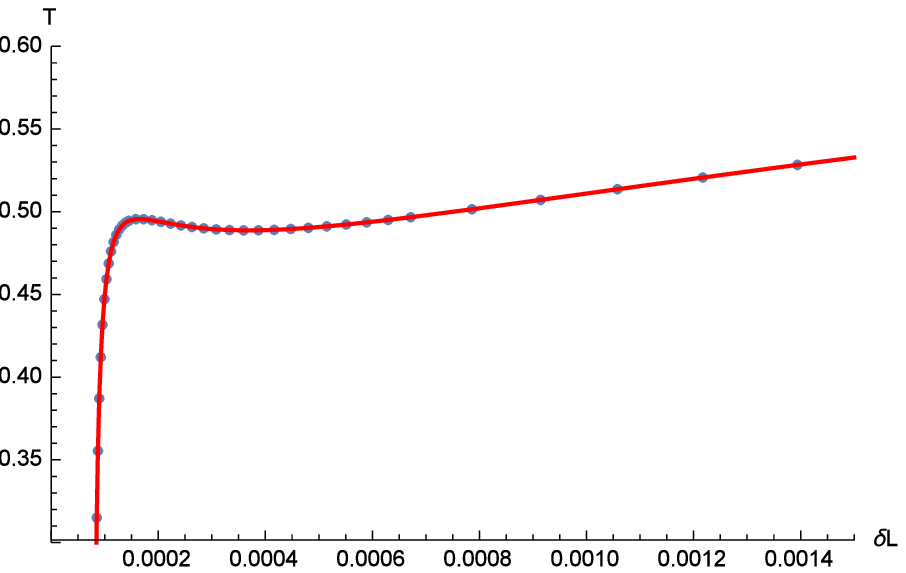}}}
\centerline{\subfigure[]{\label{2c}
\includegraphics[width=8cm,height=6cm]{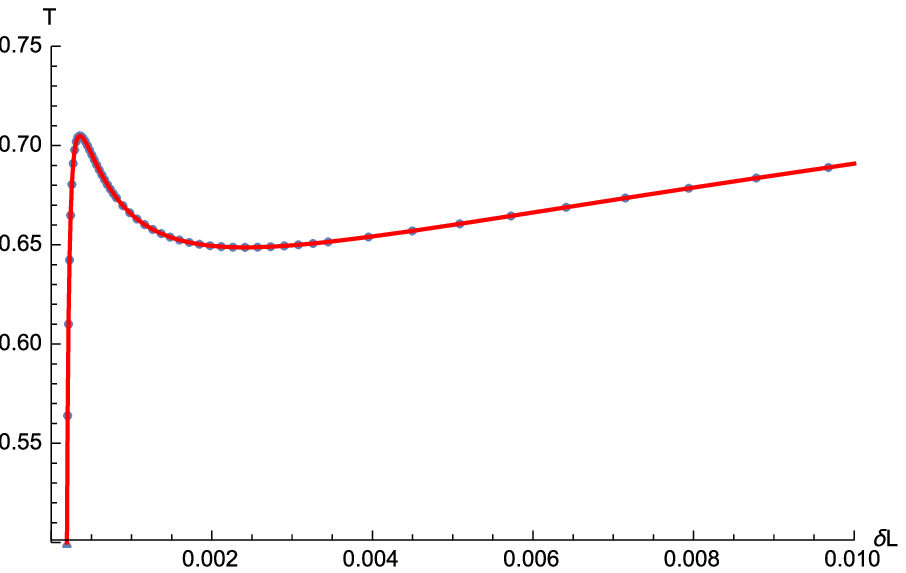}}
\subfigure[]{\label{2d}
\includegraphics[width=8cm,height=6cm]{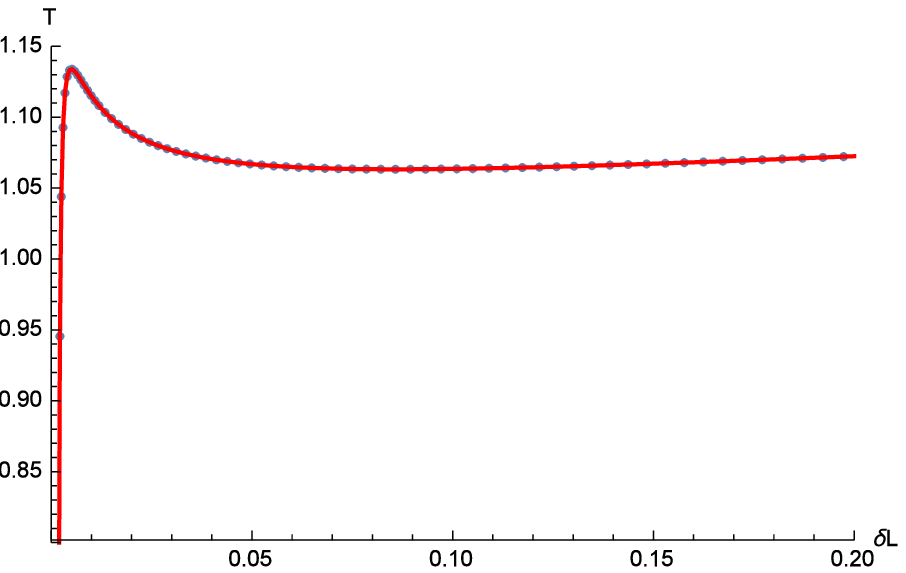}}}
 \caption{$T$ vs. $\delta L$ for $n=4, \theta_0=0.2$ (a) $\alpha=0, q=0.12$ (b) $\alpha=0.25, q=0.12$ (c) $\alpha=0.5, q=0.06$ (d) $\alpha=0.75, q=0.06$} \label{fg2}
\end{figure*}

\begin{figure*}
\centerline{\subfigure[]{\label{3a}
\includegraphics[width=8cm,height=6cm]{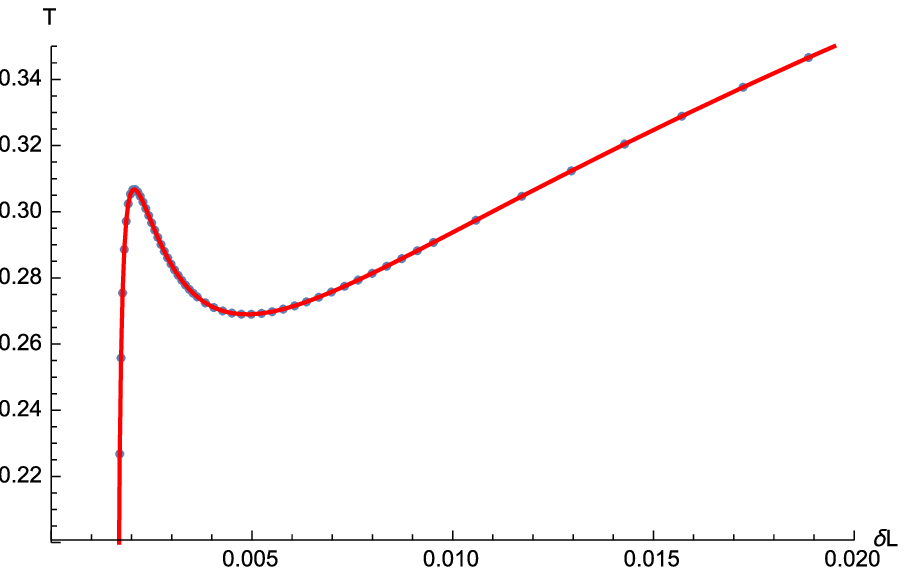}}
\subfigure[]{\label{3b}
\includegraphics[width=8cm,height=6cm]{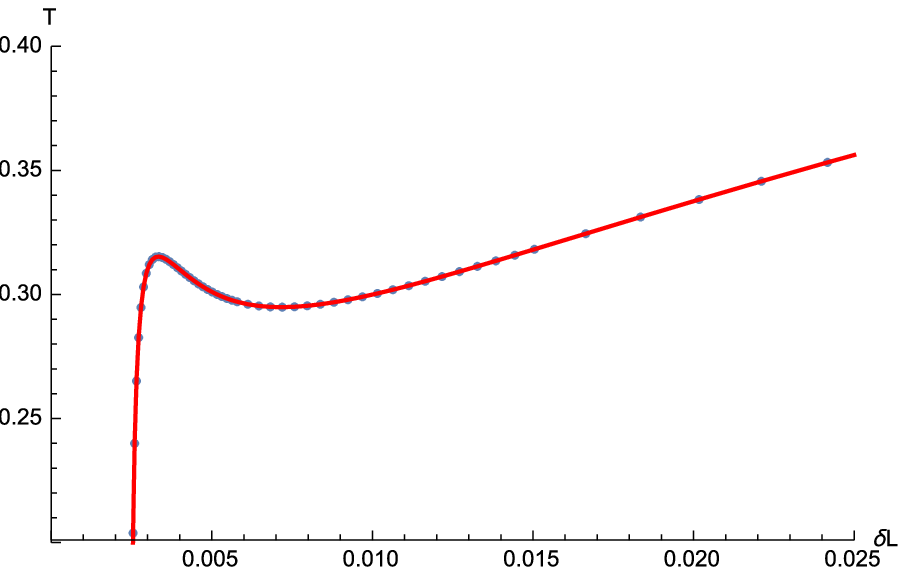}}}
\centerline{\subfigure[]{\label{3c}
\includegraphics[width=8cm,height=6cm]{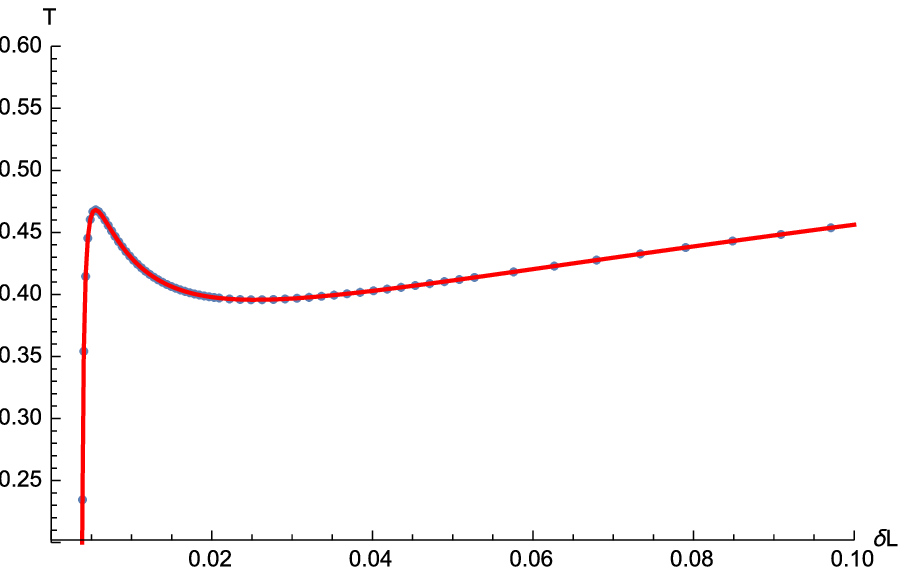}}
\subfigure[]{\label{3d}
\includegraphics[width=8cm,height=6cm]{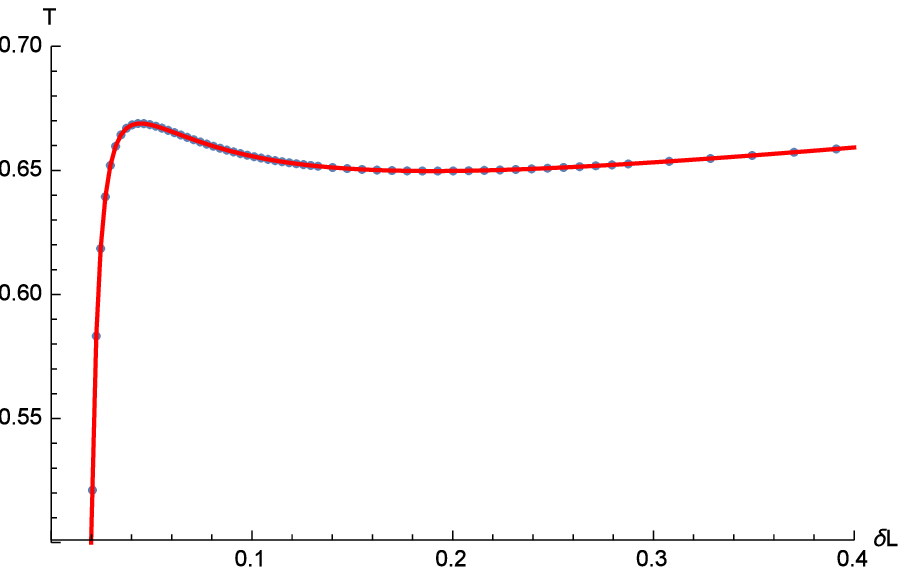}}}
 \caption{$T$ vs. $\delta L$ for $n=3, \theta_0=0.3$ (a) $\alpha=0, q=0.12$ (b) $\alpha=0.25, q=0.12$ (c) $\alpha=0.5, q=0.06$ (d) $\alpha=0.75, q=0.06$} \label{fg3}
\end{figure*}

\begin{figure*}
\centerline{\subfigure[]{\label{4a}
\includegraphics[width=8cm,height=6cm]{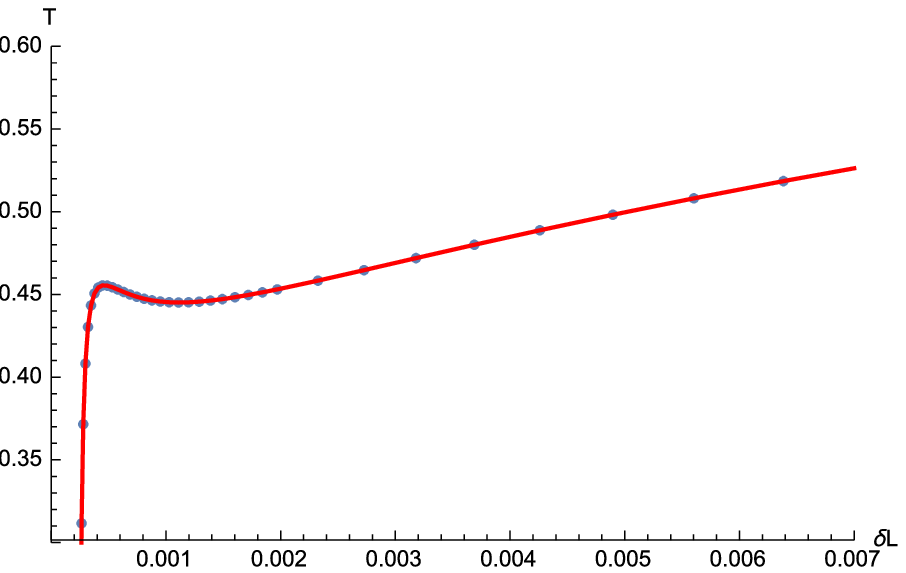}}
\subfigure[]{\label{4b}
\includegraphics[width=8cm,height=6cm]{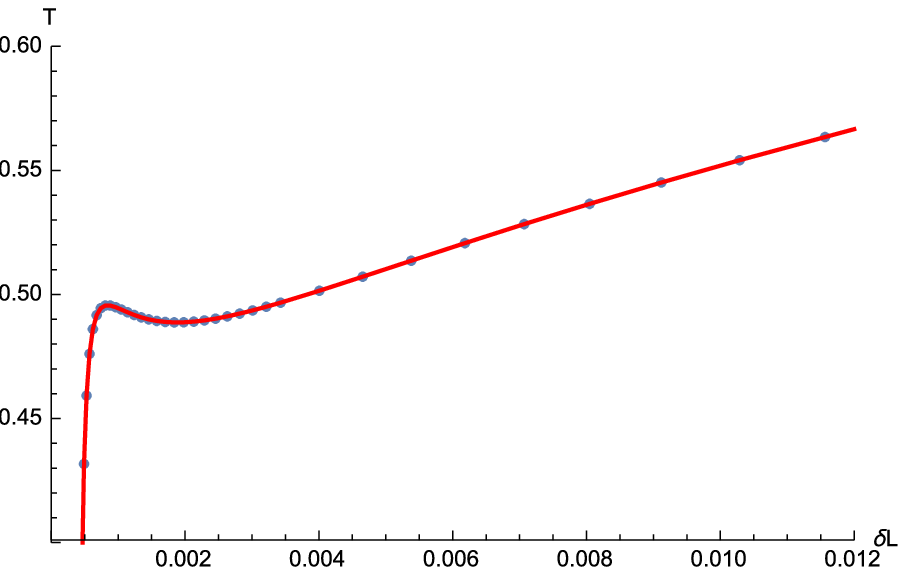}}}
\centerline{\subfigure[]{\label{4c}
\includegraphics[width=8cm,height=6cm]{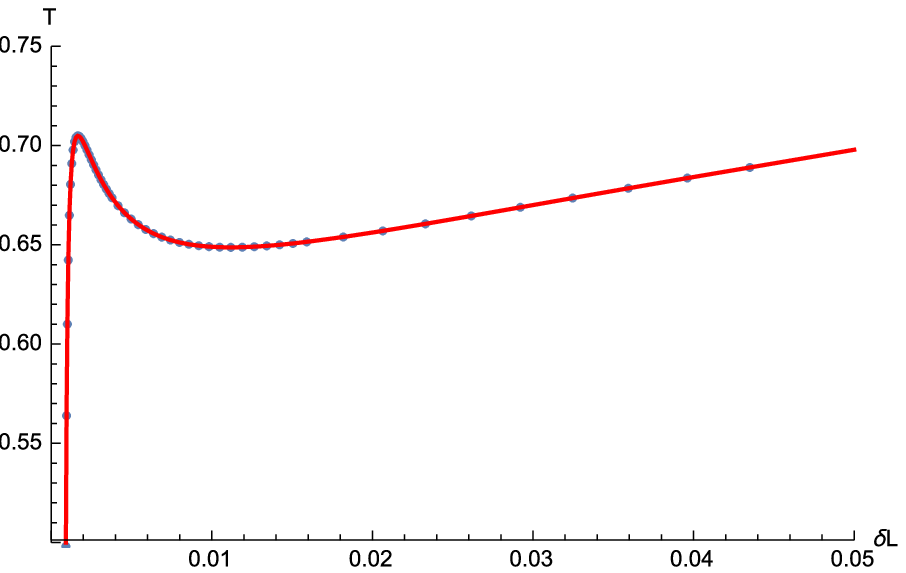}}
\subfigure[]{\label{4d}
\includegraphics[width=8cm,height=6cm]{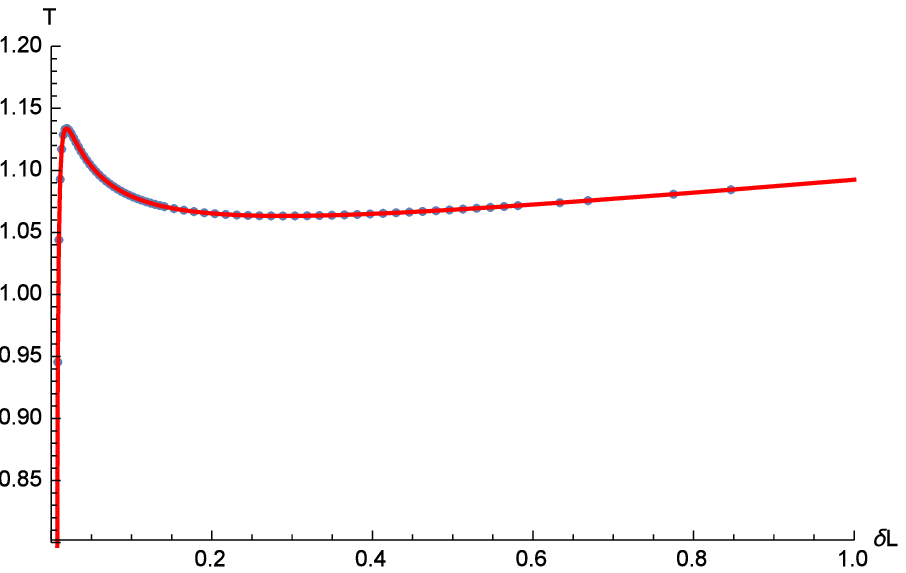}}}
 \caption{$T$ vs. $\delta L$ for $n=4, \theta_0=0.3$ (a) $\alpha=0, q=0.12$ (b) $\alpha=0.25, q=0.12$ (c) $\alpha=0.5, q=0.06$ (d) $\alpha=0.75, q=0.06$} \label{fg4}
\end{figure*}

To summarize, we consider $4\times2\times2=16$ cases due to different choices of parameters. In each case we concentrate on the case $q<q_c$ ($q_c$ denotes the critical value of the parameter $q$) to probe the possible van der Waals behavior in the $T-\delta L$ graphs. Fig. \ref{1a}-\ref{1d} show respectively the cases of $n=3, \theta_0=0.2$ with four different choices of $\alpha$ while Fig. \ref{2a}-\ref{2d} display the cases for $n=4, \theta_0=0.2$. The cases for $n=3, \theta_0=0.3$ are depicted in Fig. \ref{3a}-\ref{3d} while the cases for $n=4, \theta_0=0.3$ are depicted in Fig. \ref{4a}-\ref{4d}.

From all the $T-\delta L$ graphs for $q<q_c$, the van der Waals like behavior can be clearly witnessed. There exist both the local maximum temperature and the local minimum temperature, which we denote as $T_{max}$ and $T_{min}$ respectively. As shown in Fig. \ref{fg1}-Fig. \ref{fg4}, the effects of dilaton gravity are reflected in two ways. Firstly, as the increasing of the parameter $\alpha$, $T_{max}$ and $T_{min}$ both increase. Secondly, the corresponding $\delta L$ also increase when $\alpha$ increases. And this phenomena is more apparent for large $\alpha$. Comparing Fig. \ref{fg1} with Fig. \ref{fg2}, one may find that the case $n=4$ has higher $T_{max}$ and $T_{min}$ than the case $n=3$. This can also be witnessed by comparing Fig. \ref{fg3} with Fig. \ref{fg4}. On the other hand, the effect of boundary region size is quite apparent in the range of $\delta L$ axis when one compares Fig. \ref{fg1} with Fig. \ref{fg3}, or compares Fig. \ref{fg2} with Fig. \ref{fg4}.

\section{Numerical check of equal area law in $T-\delta L$ graph}
\label{sec:4}

According to the analogous specific heat for $T-\delta L$ graph introduced in Ref. \cite{zengxiaoxiong3}
\begin{equation}
C=T\frac{\partial \delta L}{\partial T}, \label{14}
 \end{equation}
$T-\delta L$ graph for $q<q_c$ can be divided into three branches. Namely, the stable large radius branch with positive specific heat, the unstable medium radius branch with negative specific heat and the stable small radius branch with positive specific heat. This phenomena is quite similar to that of $T-S$ graph and $P-v$ graph of dilaton AdS black holes.

One can remove the unstable branch in $T-\delta L$ curve with a bar $T=T_*$ vertical to the temperature axis by mimicking the approach of $T-S$ graph \cite{Spallucci}. Note that $T_*$ should be interpreted physically as the first order phase transition temperature and can be determined utilizing the free energy analysis. The analogous equal area law for the $T-\delta L$ graph reads
\begin{equation}
T_*\times(\delta L_3-\delta L_1)=\int^{\delta L_3}_{\delta L_1}Td\delta L,\label{15}
\end{equation}%
where $\delta L_1$, $\delta L_2$, $\delta L_3$ are three values of $\delta L$ corresponding to $T_*$. Here, we have assumed that $\delta L_1<\delta L_2<\delta L_3$. If one can prove that the L. H. S of Eq. \ref{15} equal to its R. H. S, one can draw the conclusion that the analogous equal area law holds for the $T-\delta L$ graph of dilaton AdS black holes.

\begin{figure*}
\centerline{\subfigure[]{\label{5a}
\includegraphics[width=8cm,height=6cm]{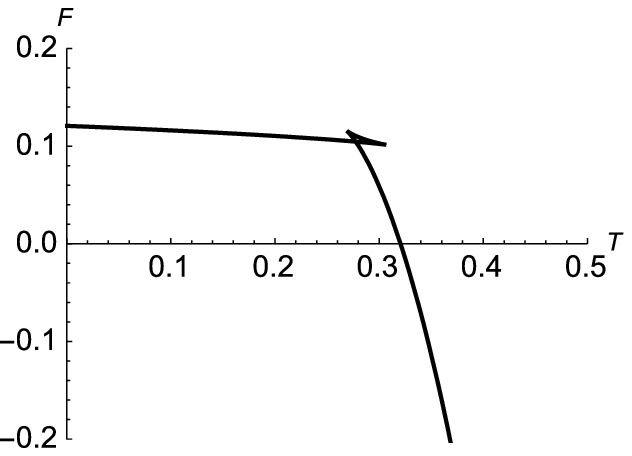}}
\subfigure[]{\label{5b}
\includegraphics[width=8cm,height=6cm]{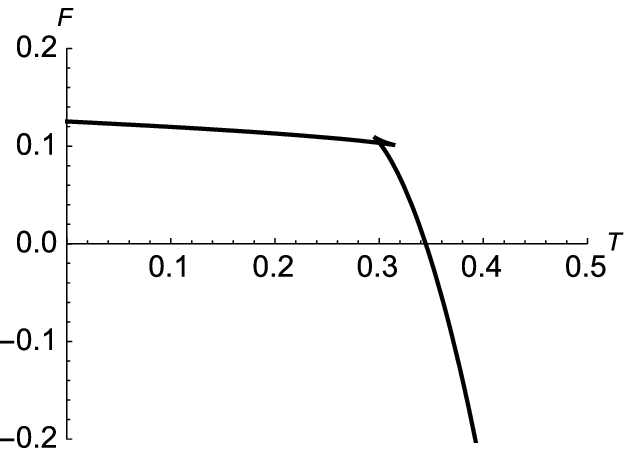}}}
\centerline{\subfigure[]{\label{5c}
\includegraphics[width=8cm,height=6cm]{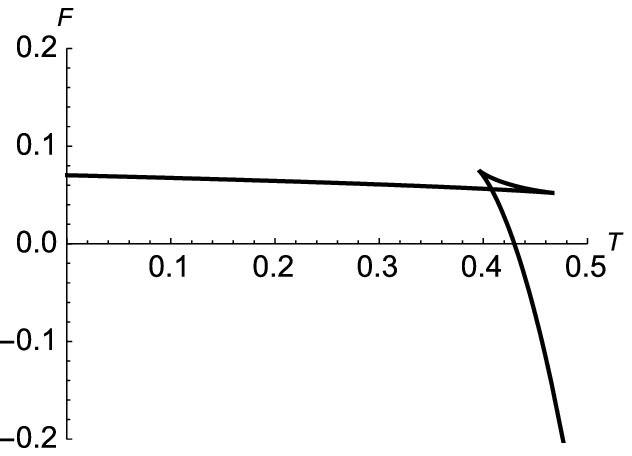}}
\subfigure[]{\label{5d}
\includegraphics[width=8cm,height=6cm]{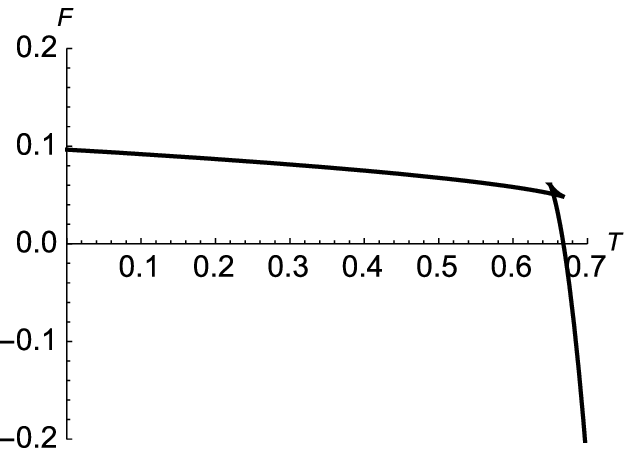}}}
 \caption{$F$ vs. $T$ for $n=3, l=1, b=1, k=1$ (a) $\alpha=0,q=0.12$ (b) $\alpha=0.25,q=0.12$ (c) $\alpha=0.50,q=0.06$ (d) $\alpha=0.75,q=0.06$} \label{fg5}
\end{figure*}

\begin{figure*}
\centerline{\subfigure[]{\label{6a}
\includegraphics[width=8cm,height=6cm]{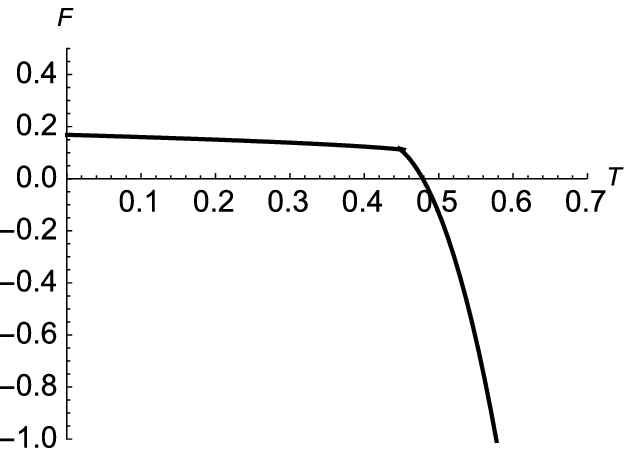}}
\subfigure[]{\label{6b}
\includegraphics[width=8cm,height=6cm]{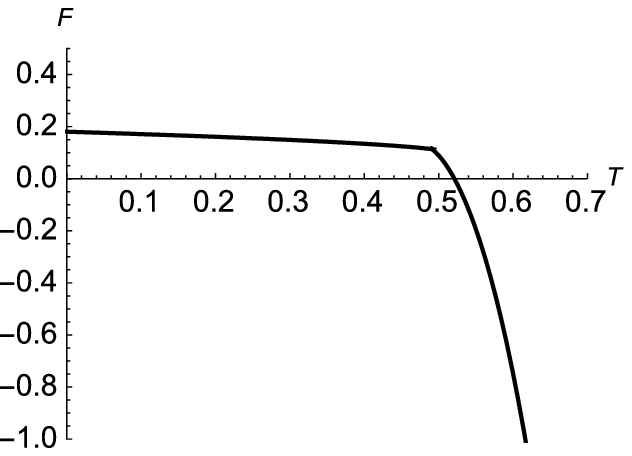}}}
\centerline{\subfigure[]{\label{6c}
\includegraphics[width=8cm,height=6cm]{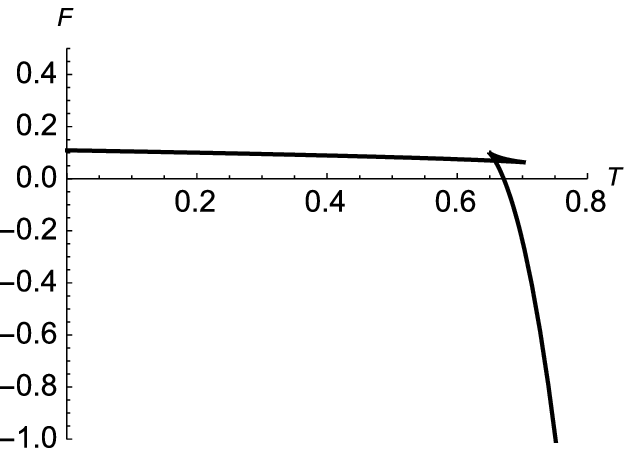}}
\subfigure[]{\label{6d}
\includegraphics[width=8cm,height=6cm]{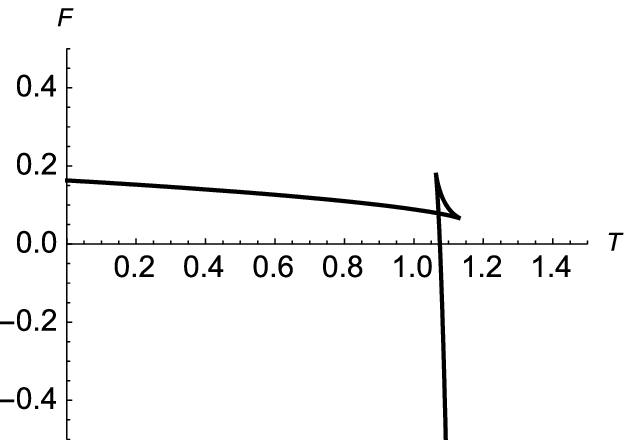}}}
 \caption{$F$ vs. $T$ for $n=4, l=1, b=1, k=1$ (a) $\alpha=0,q=0.12$ (b) $\alpha=0.25,q=0.12$ (c) $\alpha=0.50,q=0.06$ (d) $\alpha=0.75,q=0.06$} \label{fg6}
\end{figure*}

Before we carry out the examination, it is urgent to study the behavior of free energy first. Utilizing Eqs. \ref{4}, \ref{6}, \ref{9} and \ref{10}, the free energy of dilaton AdS black holes can be derived as
\begin{eqnarray}
F&=&M-TS=\frac{\omega_{n-1}(1+\alpha^2)b^{(n-1)\gamma}r_+^{n-2+\gamma-n\gamma}}{16\pi}\times \Big\{\frac{k(n-2)r_+^{2\gamma}b^{-2\gamma}}{n-2+\alpha^2}+\frac{nb^{2\gamma}r_+^{2-2\gamma}(\alpha^2-1)}{l^2(n-\alpha^2)}
\nonumber
\\
&\;&+\frac{2b^{-2(n-2)\gamma}q^2r_+^{2(n-2)(\gamma-1)}(2n-3+\alpha^2)}{(n-1)(n-2+\alpha^2)}\Big\},\label{16}
\end{eqnarray}%
with the $F-T$ graphs depicted in Fig. \ref{fg5} and Fig. \ref{fg6} for different choices of parameters.

One can observe the familiar swallow tail behavior in all the $F-T$ graphs and obtain $T_*$ from the intersection point of two branches in the graphs. Comparing these graphs, one may conclude that the first order phase transition temperature $T_*$ is affected by both $\alpha$ and $n$. When $\alpha$ increases, $T_*$ increases. When $n$ increases, $T_*$ increases too. With $T_*$ at hand, one can calculate the left-hand side and right-hand side of Eq.(\ref{15}) for different cases. As can be seen from Table \ref{tb1}-Table \ref{tb4}, the relative errors for all the cases are small enough and we can safely conclude that the analogous equal area law holds for $T-\delta L$ graph of dilaton AdS black holes.

\begin{table}[!h]
\tabcolsep 0pt
\caption{Numerical check of equal area law in $T-\delta L$ graph for $n=3, \theta_0=0.2, l=1, b=1, k=1$}
\vspace*{-12pt}
\begin{center}
\def\temptablewidth{1\textwidth}
{\rule{\temptablewidth}{2pt}}
\begin{tabular*}{\temptablewidth}{@{\extracolsep{\fill}}c||c||c||c||c||c||c||c}
 $\alpha$ & $q$ &  $T_*$ & $\delta L_1$  &$\delta L_3$ &$T_*(\delta L_3-\delta L_1)$ & $\int^{\delta L_3}_{\delta L_1}Td\delta L$ & relative error \\   \hline
0      & 0.12          &0.277496     & 0.0000637680      & 0.0002613560 & 0.0000548299 & 0.0000549115& $0.1486 \%$ \\
 0.25      & 0.12          &0.3002889      & 0.0007501943    & 0.0030534139 & 0.0006916313 & 0.0006854523 & $0.9014\%$ \\
 0.50      & 0.06          &0.4078042      & 0.0014369702       & 0.0157774345 & 0.0058481016 & 0.0058594007 & $0.1928 \%$ \\
 0.75      & 0.06          &0.65401625      & 0.0130252955       & 0.1416948822 & 0.0841520006 & 0.0841644278 & $0.0148\%$
       \end{tabular*}
       {\rule{\temptablewidth}{2pt}}
       \end{center}
       \label{tb1}
       \end{table}

\begin{table}[!h]
\tabcolsep 0pt
\caption{Numerical check of equal area law in $T-\delta L$ graph for $n=4, \theta_0=0.2, l=1, b=1, k=1$}
\vspace*{-12pt}
\begin{center}
\def\temptablewidth{1\textwidth}
{\rule{\temptablewidth}{2pt}}
\begin{tabular*}{\temptablewidth}{@{\extracolsep{\fill}}c||c||c||c||c||c||c||c}
 $\alpha$ & $q$ & $T_*$ & $\delta L_1$  &$\delta L_3$ &$T_*(\delta L_3-\delta L_1)$ & $\int^{\delta L_3}_{\delta L_1}Td\delta L$ & relative error \\   \hline
 0      & 0.12        &0.448178      &0.0000688919     & 0.0003011692 & 0.0001041016 & 0.0001041083 & $0.0064 \%$ \\
 0.25      & 0.12          &0.490898     & 0.0001308121      & 0.0005050468 & 0.0001837111 & 0.0001837165 & $0.0029\%$ \\
 0.50     & 0.06       &0.65793667      & 0.0002371822     & 0.0046525001 & 0.0029049996 & 0.0029063836 & $0.0476 \%$ \\
0.75      & 0.06          &1.070886315     & 0.0027446929     & 0.1854763902 & 0.1956848740 & 0.1957787303 & $0.0479\%$
       \end{tabular*}
       {\rule{\temptablewidth}{2pt}}
       \end{center}
       \label{tb2}
       \end{table}

 \begin{table}[!h]
\tabcolsep 0pt
\caption{Numerical check of equal area law in $T-\delta L$ graph for $n=3, \theta_0=0.3, l=1, b=1, k=1$}
\vspace*{-12pt}
\begin{center}
\def\temptablewidth{1\textwidth}
{\rule{\temptablewidth}{2pt}}
\begin{tabular*}{\temptablewidth}{@{\extracolsep{\fill}}c||c||c||c||c||c||c||c}
 $\alpha$ & $q$ &  $T_*$ & $\delta L_1$  &$\delta L_3$ &$T_*(\delta L_3-\delta L_1)$ & $\int^{\delta L_3}_{\delta L_1}Td\delta L$ & relative error \\   \hline
0      & 0.12          &0.277496     & 0.0017837032      & 0.0073122615 & 0.0015341528 & 0.0015366458& $0.1622 \%$ \\
 0.25      & 0.12          &0.3002889      & 0.0024892165   & 0.0101203310 & 0.0022915390 & 0.0022710854 & $0.9006\%$ \\
 0.50      & 0.06          &0.4078042      & 0.0042562477     & 0.0460392623 & 0.0170392888 & 0.0170769726 & $0.2207 \%$ \\
 0.75      & 0.06          &0.65401625      & 0.0301340104       & 0.3145252201 & 0.1859964725 & 0.1860497056 & $0.0286\%$
       \end{tabular*}
       {\rule{\temptablewidth}{2pt}}
       \end{center}
       \label{tb3}
       \end{table}

       \begin{table}[!h]
\tabcolsep 0pt
\caption{Numerical check of equal area law in $T-\delta L$ graph for $n=4, \theta_0=0.3, l=1, b=1, k=1$}
\vspace*{-12pt}
\begin{center}
\def\temptablewidth{1\textwidth}
{\rule{\temptablewidth}{2pt}}
\begin{tabular*}{\temptablewidth}{@{\extracolsep{\fill}}c||c||c||c||c||c||c||c}
 $\alpha$ & $q$ & $T_*$ & $\delta L_1$  &$\delta L_3$ &$T_*(\delta L_3-\delta L_1)$ & $\int^{\delta L_3}_{\delta L_1}Td\delta L$ & relative error \\   \hline
 0      & 0.12        &0.448178     & 0.0003666396     & 0.0015882967 & 0.0005475198 &0.0005475520& $0.0059\%$ \\
 0.25      & 0.12          &0.490898     & 0.0006709414     & 0.0025817161 & 0.0009379955 & 0.0009380156 & $0.0021\%$ \\
 0.50     & 0.06       &0.65793667      & 0.0011076791    & 0.0213463021 & 0.0133157322 & 0.0133239547 & $0.0617 \%$ \\
0.75      & 0.06          &1.070886315     & 0.0104051412     & 0.5641278672 & 0.5929740896 & 0.5938277878 & $0.1438\%$
       \end{tabular*}
       {\rule{\temptablewidth}{2pt}}
       \end{center}
       \label{tb4}
       \end{table}

\section{Entanglement entropy of dilaton AdS black holes and its van der Waals like behavior}
\label{sec:5}
Besides the two point correlation function, entanglement entropy serves as another nonlocal observable to probe the holographic properties of black holes.

The entanglement entropy can be expressed holographically in terms of the area of a minimal surface anchored on $\partial A$ as \cite{Takayanagi1,Takayanagi2}
 \begin{equation}
S_A=\frac{Area(\Sigma)}{4G_N},\label{l7}
\end{equation}
where $\Sigma$ is the codimension-2 minimal surface with boundary condition $\partial \Sigma=\partial A$ and $G_N$ is the Newton's constant.

As argued in Ref. \cite{Nguyen}, one can avoid dealing with the phase transition between connected and disconnected minimal surfaces by avoiding large entangling regions. For this consideration, we choose the region to be a spherical cap on the boundary delimited by $\theta\leq\theta_0$.

Parametrizing the minimal surface by the function $r(\theta)$, the holographic entanglement entropy can be obtained as
 \begin{equation}
S_A=\frac{\pi}{2}\int^{\theta_0}_0 r\sin \theta \sqrt{\frac{r'^2}{f(r)}+r^2} d \theta,\label{l8}
\end{equation}
where $r'=dr/ d\theta$. From the above equation, one can obtain 
\begin{eqnarray}
\mathcal{L}=r\sin \theta \sqrt{\frac{r'^2}{f(r)}+r^2}. \label{19}
 \end{eqnarray}
Substituting it into Euler-Lagrange equation, one can derive the equation for $r(\theta)$. Then one can obtain $r(\theta)$ by solving the equation constrained by the boundary condition $r(0)= r_0, r'(0)=0$ numerically.

To avoid the divergence, the holographic entanglement entropy should be regularized by subtracting the entanglement entropy $S_0$ ($S_0$ can also be obtained through numerical treatment) in pure AdS with the same boundary region. We denote the regularized entanglement entropy as $\delta S$ and mainly deal with the behavior of $T-\delta S$ graph in the following discussions.

As in Sec.~\ref {sec:3}, we will consider $4\times2\times2=16$ cases due to different choices of parameters. For the sake of clarity, we repeat the choices here. $\alpha$ is chosen as $0, 0.25, 0.5, 0.75$ respectively to investigate the effect of dilaton gravity on the phase structure of holographic entanglement entropy while $n$ is chosen as $3,4$ respectively to probe the effect of spacetime dimensionality. $\theta_0$ is chosen as $0.2, 0.3$ respectively to study the effect of different boundary region sizes. And the cutoff $\theta_c$ will also be chosen as $0.199, 0.299$ accordingly. In each case we focus on the case $q<q_c$ to probe the possible van der Waals behavior.

\begin{figure*}
\centerline{\subfigure[]{\label{7a}
\includegraphics[width=8cm,height=6cm]{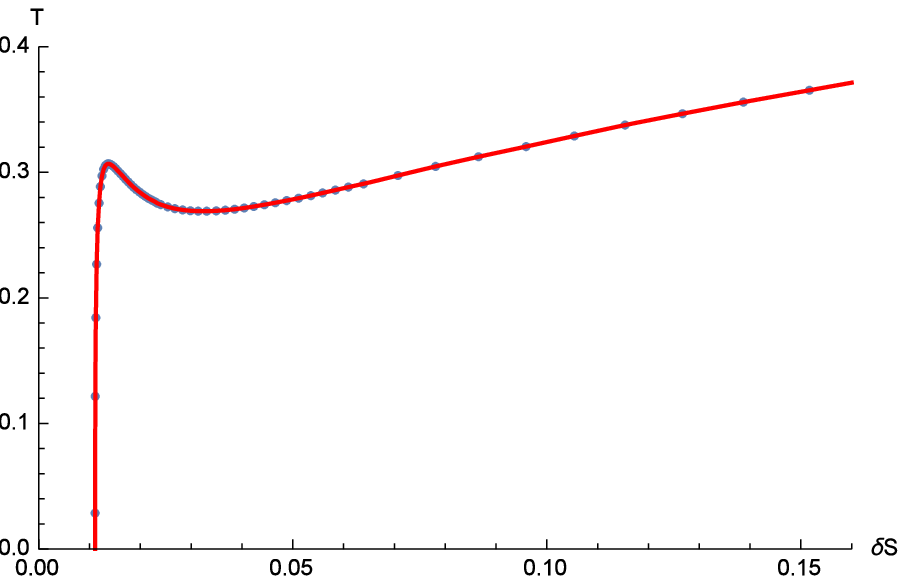}}
\subfigure[]{\label{7b}
\includegraphics[width=8cm,height=6cm]{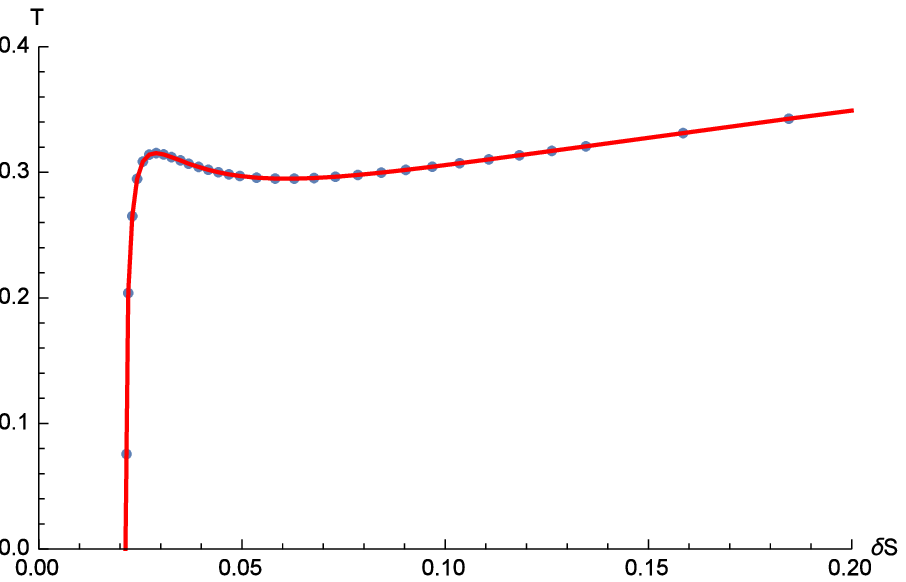}}}
\centerline{\subfigure[]{\label{7c}
\includegraphics[width=8cm,height=6cm]{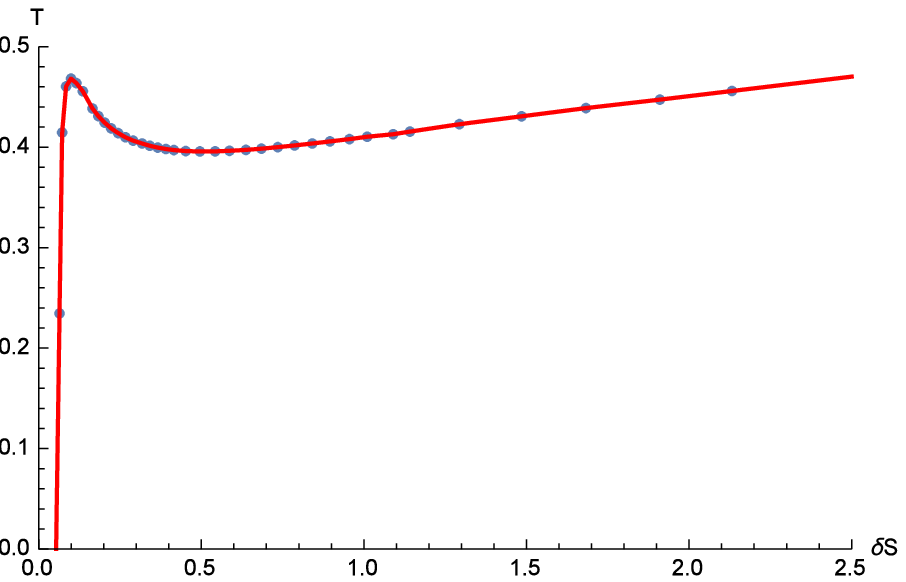}}
\subfigure[]{\label{7d}
\includegraphics[width=8cm,height=6cm]{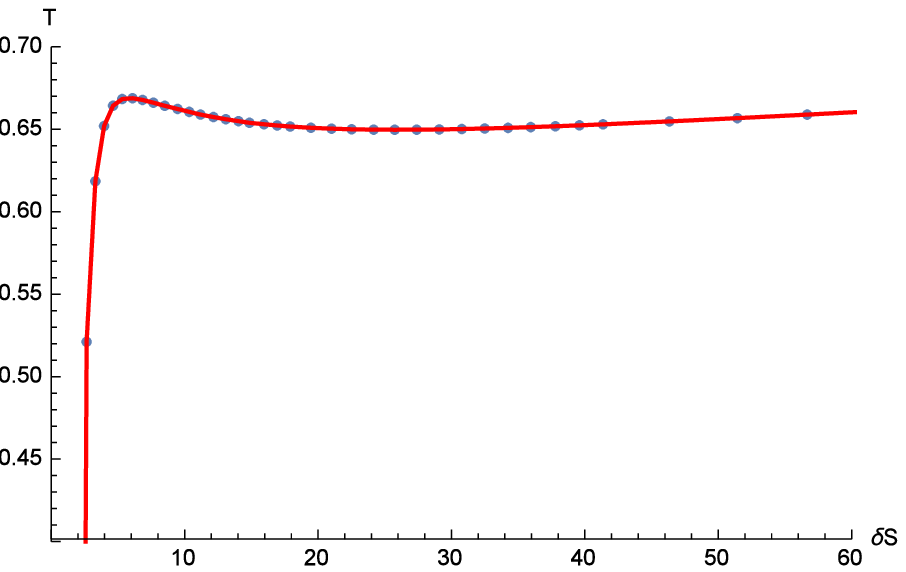}}}
 \caption{$T$ vs. $\delta S$ for $n=3, \theta_0=0.2$ (a) $\alpha=0, q=0.12$ (b) $\alpha=0.25, q=0.12$ (c) $\alpha=0.5, q=0.06$ (d) $\alpha=0.75, q=0.06$} \label{fg7}
\end{figure*}

\begin{figure*}
\centerline{\subfigure[]{\label{8a}
\includegraphics[width=8cm,height=6cm]{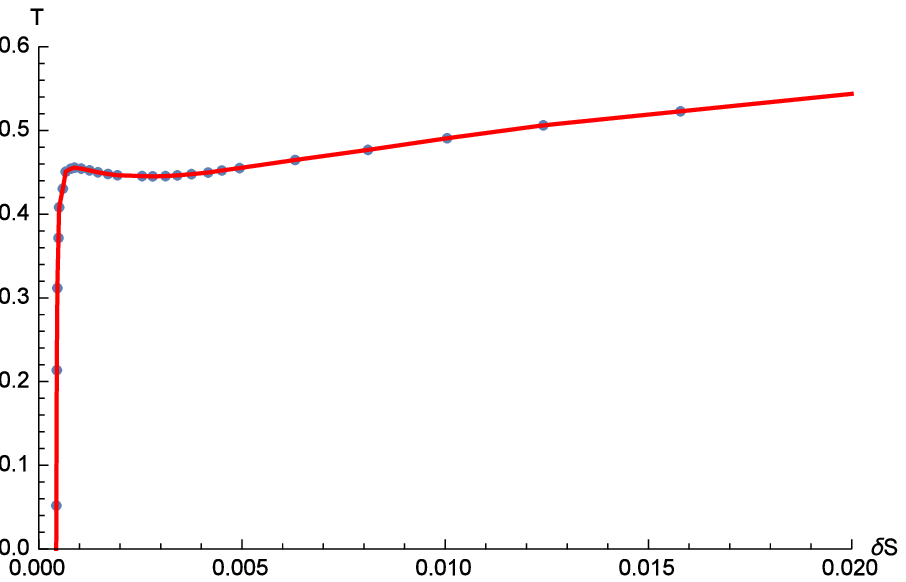}}
\subfigure[]{\label{8b}
\includegraphics[width=8cm,height=6cm]{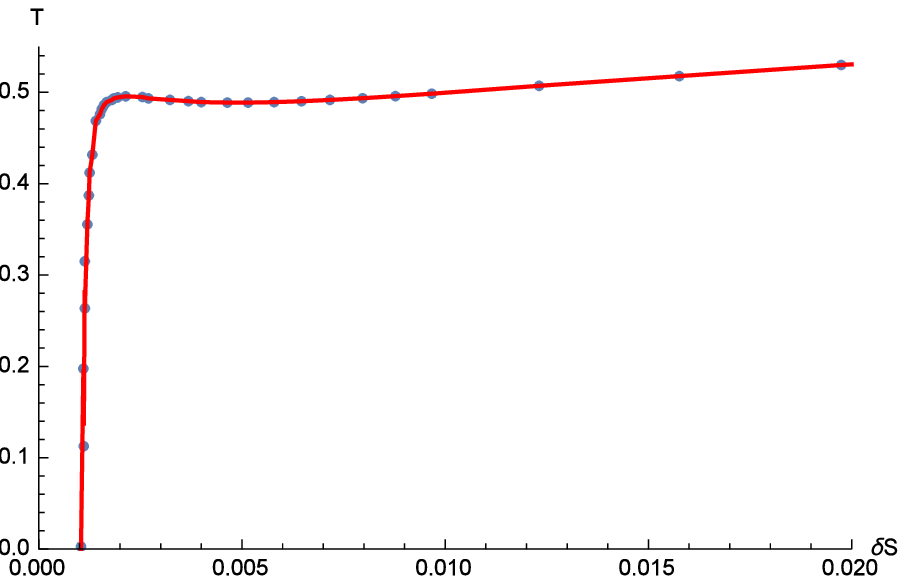}}}
\centerline{\subfigure[]{\label{8c}
\includegraphics[width=8cm,height=6cm]{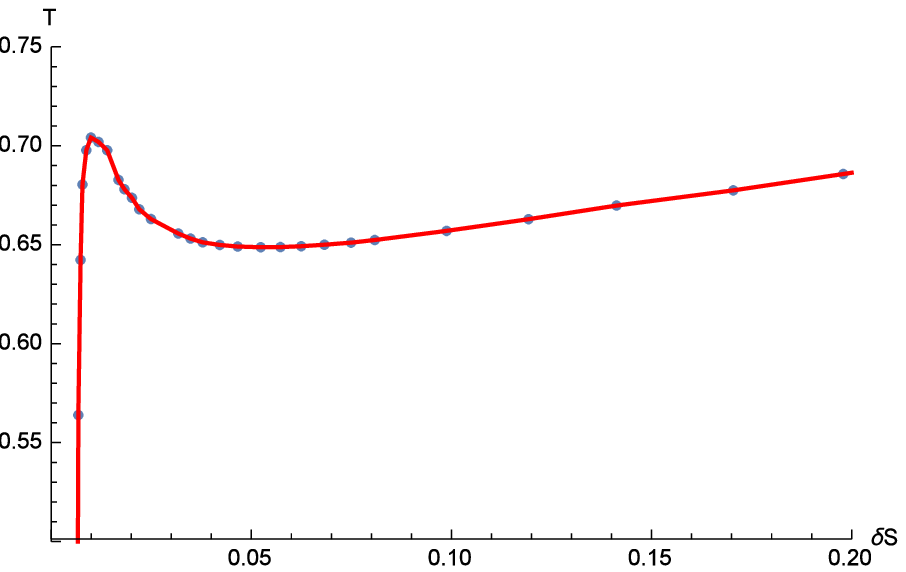}}
\subfigure[]{\label{8d}
\includegraphics[width=8cm,height=6cm]{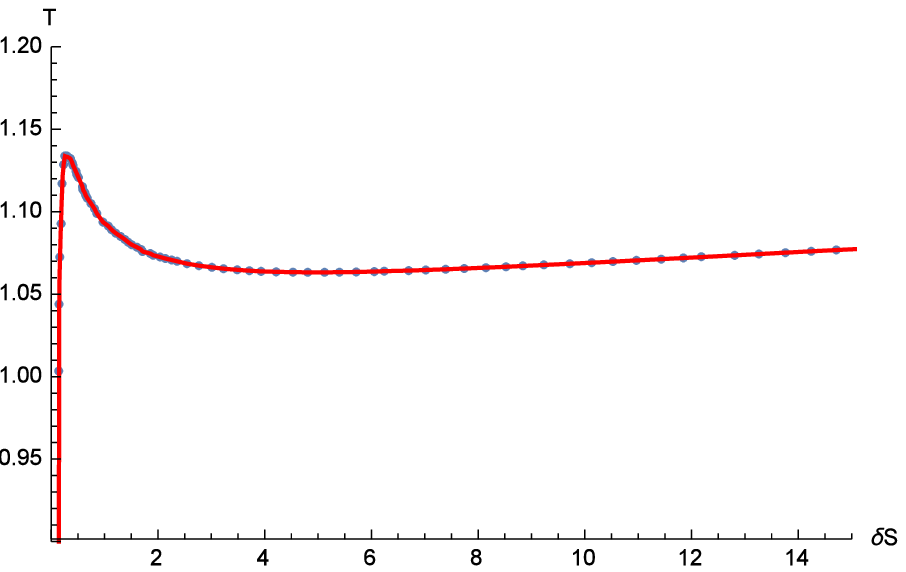}}}
 \caption{$T$ vs. $\delta S$ for $n=4, \theta_0=0.2$ (a) $\alpha=0, q=0.12$ (b) $\alpha=0.25, q=0.12$ (c) $\alpha=0.5, q=0.06$ (d) $\alpha=0.75, q=0.06$} \label{fg8}
\end{figure*}

\begin{figure*}
\centerline{\subfigure[]{\label{9a}
\includegraphics[width=8cm,height=6cm]{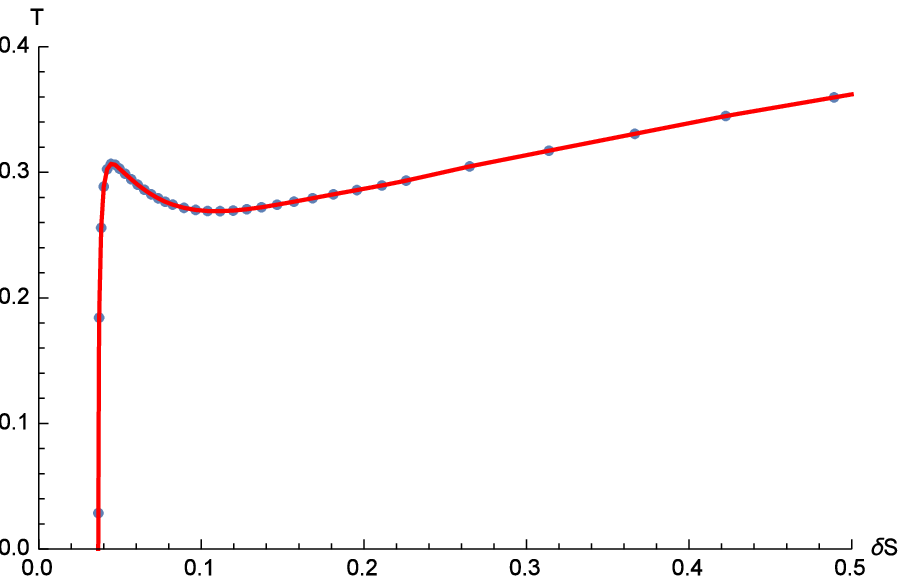}}
\subfigure[]{\label{9b}
\includegraphics[width=8cm,height=6cm]{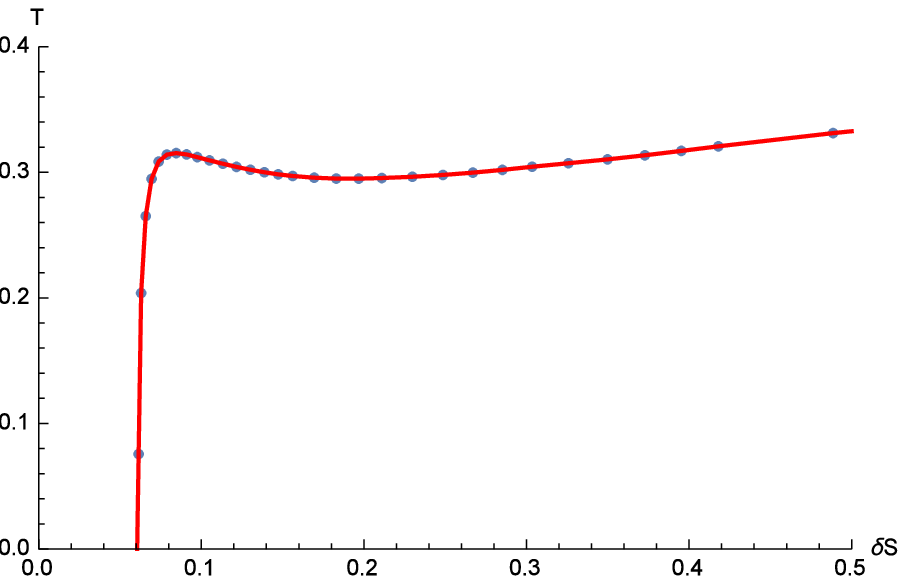}}}
\centerline{\subfigure[]{\label{9c}
\includegraphics[width=8cm,height=6cm]{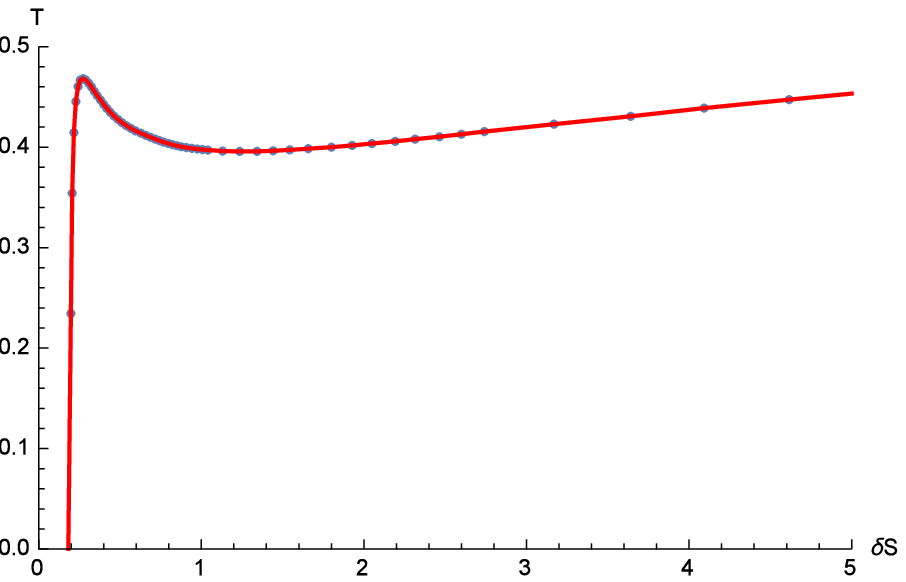}}
\subfigure[]{\label{9d}
\includegraphics[width=8cm,height=6cm]{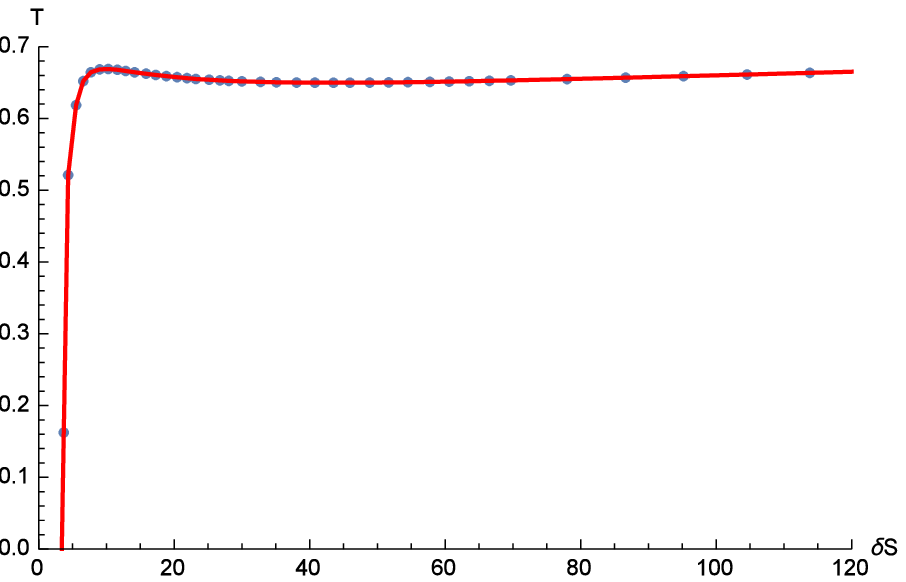}}}
 \caption{$T$ vs. $\delta S$ for $n=3, \theta_0=0.3$ (a) $\alpha=0, q=0.12$ (b) $\alpha=0.25, q=0.12$ (c) $\alpha=0.5, q=0.06$ (d) $\alpha=0.75, q=0.06$} \label{fg9}
\end{figure*}

\begin{figure*}
\centerline{\subfigure[]{\label{10a}
\includegraphics[width=8cm,height=6cm]{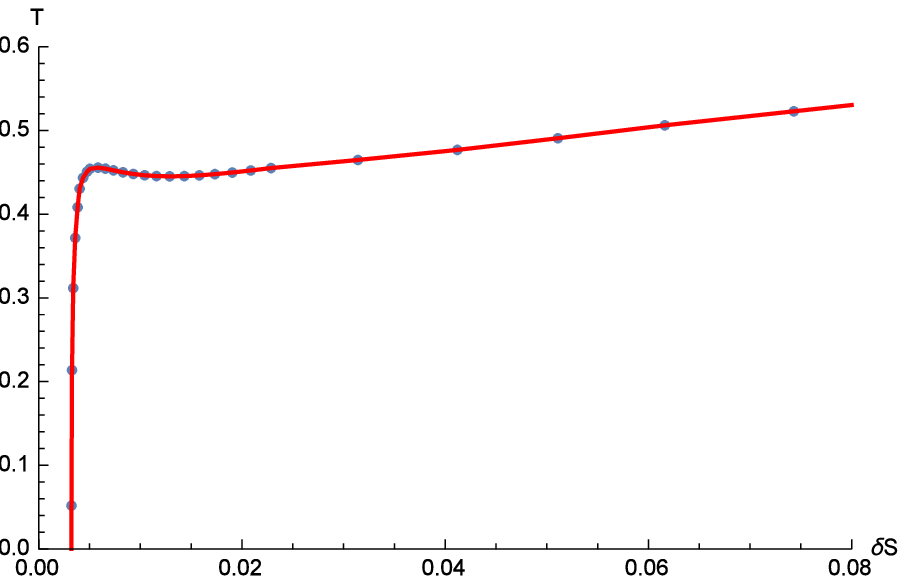}}
\subfigure[]{\label{10b}
\includegraphics[width=8cm,height=6cm]{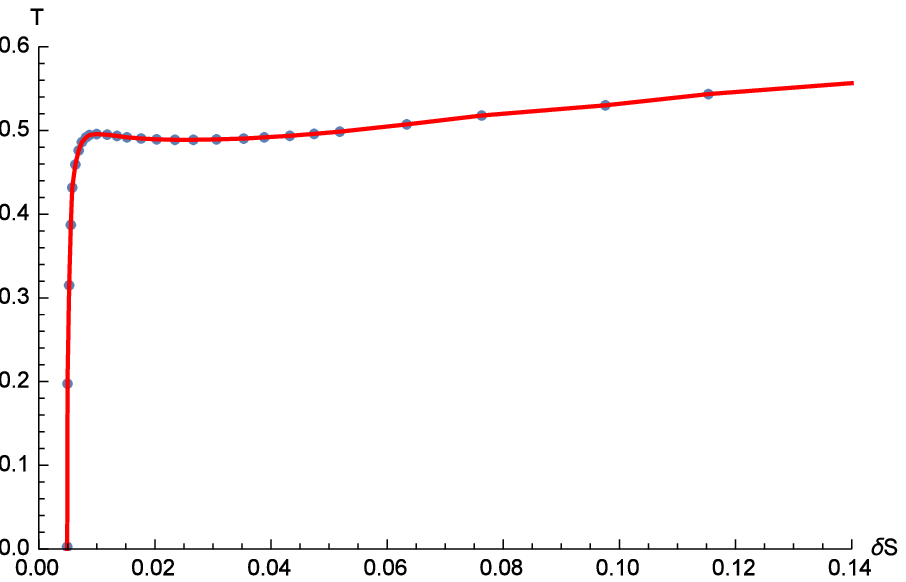}}}
\centerline{\subfigure[]{\label{10c}
\includegraphics[width=8cm,height=6cm]{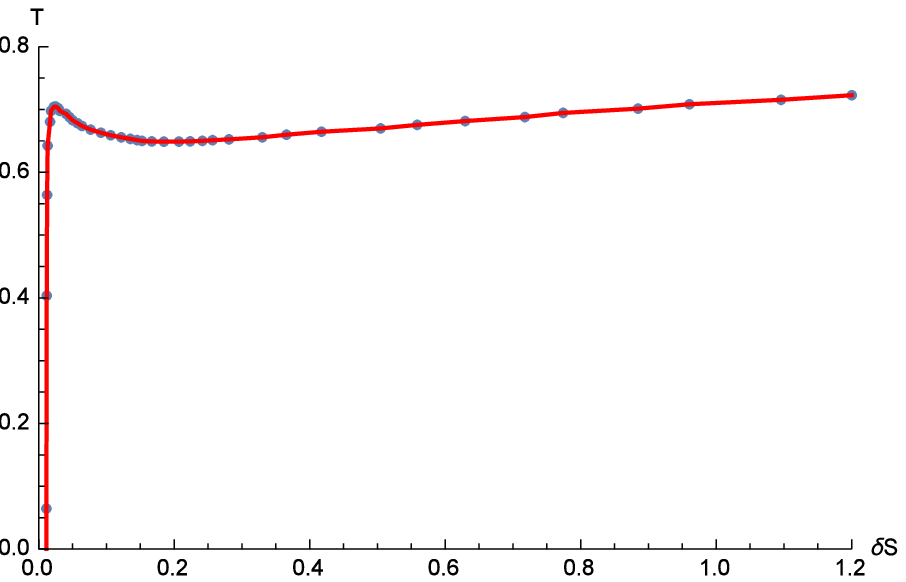}}
\subfigure[]{\label{10d}
\includegraphics[width=8cm,height=6cm]{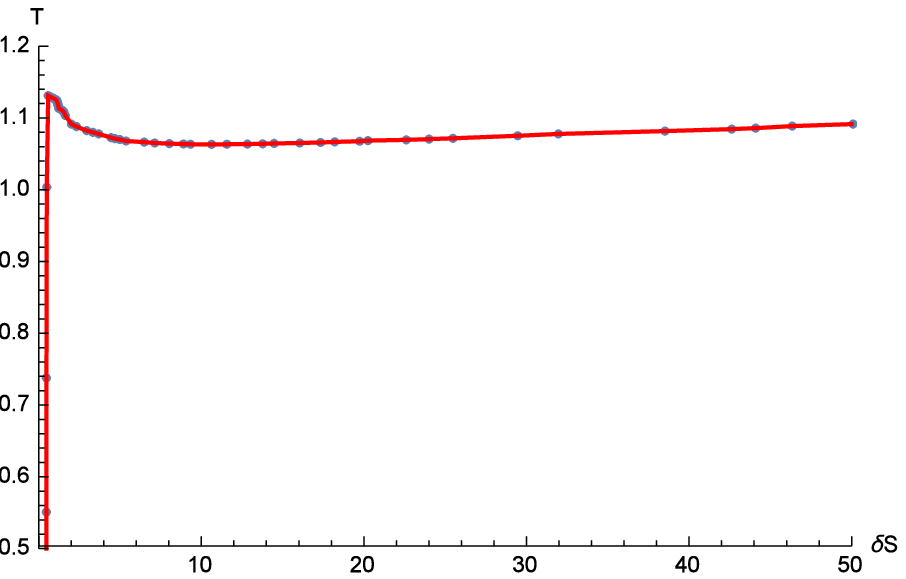}}}
 \caption{$T$ vs. $\delta S$ for $n=4, \theta_0=0.3$ (a) $\alpha=0, q=0.12$ (b) $\alpha=0.25, q=0.12$ (c) $\alpha=0.5, q=0.06$ (d) $\alpha=0.75, q=0.06$} \label{fg10}
\end{figure*}

We plot the cases of $n=3, \theta_0=0.2$ in Fig. \ref{7a}-\ref{7d} and the cases of  $n=4, \theta_0=0.2$ in Fig. \ref{8a}-\ref{8d}. The cases for $n=3, \theta_0=0.3$ are displayed in Fig. \ref{9a}-\ref{9d} while the cases for $n=4, \theta_0=0.3$ are depicted in Fig. \ref{10a}-\ref{10d}.

The van der Waals like behavior can be clearly witnessed from all the $T-\delta S$ graphs for $q<q_c$. As the $T-\delta L$ graphs presented in Sec.~\ref {sec:3}, there also exist both the local maximum temperature and the local minimum temperature in all the $T-\delta S$ graphs. As depicted in  Fig. \ref{fg7}-Fig. \ref{fg10}, not only $T_{max}$ and $T_{min}$ but also the corresponding $\delta S$ increase as the increasing of the parameter $\alpha$, showing the impact of dilaton gravity. Comparing Fig. \ref{fg7} with Fig. \ref{fg8}, or comparing Fig. \ref{fg9} with Fig. \ref{fg10}, it is not difficult to observe that the cases $n=4$ have higher $T_{max}$ and $T_{min}$ than the cases $n=3$. Moreover, the range of $\delta S$ axis for the cases $\theta_0=0.2$ differs from that of the cases $\theta_0=0.3$, showing the effect of boundary region size. These phenomena are quite similar to those of two point correlation function probed in Sec.~\ref {sec:3}.

\section{Numerical check of equal area law in $T-\delta S$ graph}
\label{sec:6}

The analogous specific heat for $T-\delta S$ graph has been defined in former literature \cite{zengxiaoxiong3} as
\begin{equation}
C=T\frac{\partial \delta S}{\partial T}. \label{20}
 \end{equation}
Based on the above definition, one can divide each $T-\delta S$ graph for $q<q_c$ into three branches. Namely, the stable large radius branch with positive specific heat, the unstable medium radius branch with negative specific heat and the stable small radius branch with positive specific heat. This phenomena is quite similar to that of $T-S$ graph and $P-v$ graph of dilaton AdS black holes. It is also similar to that of $T-\delta L$ graph discussed in Sec.~\ref {sec:3}.

Taking a similar treatment, one can remove the unstable branch in $T-\delta S$ curve with a bar $T=T_*$ vertical to the temperature axis. And the analogous equal area law for the $T-\delta S$ graph can be written as
\begin{equation}
T_*\times(\delta S_3-\delta S_1)=\int^{\delta S_3}_{\delta S_1}Td\delta S,\label{21}
\end{equation}%
where $\delta S_1$, $\delta S_2$, $\delta S_3$ are three values of $\delta S$ corresponding to $T_*$ with the assumption that $\delta S_1<\delta S_2<\delta S_3$.

We calculate the left-hand side and right-hand side of Eq.(\ref{21}) for different cases and list the results in Table \ref{tb5}- Table \ref{tb8}, from which one can see clearly that the relative errors for all the cases are amazingly small. So we can safely draw the conclusion that the analogous equal area law  Eq.(\ref{21}) holds for $T-\delta S$ graph of dilaton AdS black holes.

\begin{table}[!h]
\tabcolsep 0pt
\caption{Numerical check of equal area law in $T-\delta S$ graph for $n=3, \theta_0=0.2, l=1, b=1, k=1$}
\vspace*{-12pt}
\begin{center}
\def\temptablewidth{1\textwidth}
{\rule{\temptablewidth}{2pt}}
\begin{tabular*}{\temptablewidth}{@{\extracolsep{\fill}}c||c||c||c||c||c||c||c}
 $\alpha$ & $q$ &  $T_*$ & $\delta S_1$  &$\delta S_3$ &$T_*(\delta S_3-\delta S_1)$ & $\int^{\delta S_3}_{\delta S_1}Td\delta S$ & relative error \\   \hline
0      & 0.12          &0.277496     & 0.0119147123      & 0.0488420932 & 0.0102472005 & 0.0102600904& $0.1256 \%$ \\
 0.25      & 0.12          &0.3002889      & 0.0247660099    & 0.0858928910 & 0.0183557239 & 0.0183641401 & $0.0458\%$ \\
 0.50      & 0.06          &0.4078042      & 0.0716592283      & 0.9511787777 & 0.3586717662 & 0.3592798682 & $0.1693 \%$ \\
 0.75      & 0.06          &0.65401625      & 4.0720587420    & 44.1845284093 & 26.2342069900 & 26.2343780610 & $0.0007\%$
       \end{tabular*}
       {\rule{\temptablewidth}{2pt}}
       \end{center}
       \label{tb5}
       \end{table}

\begin{table}[!h]
\tabcolsep 0pt
\caption{Numerical check of equal area law in $T-\delta S$ graph for $n=4, \theta_0=0.2, l=1, b=1, k=1$}
\vspace*{-12pt}
\begin{center}
\def\temptablewidth{1\textwidth}
{\rule{\temptablewidth}{2pt}}
\begin{tabular*}{\temptablewidth}{@{\extracolsep{\fill}}c||c||c||c||c||c||c||c}
 $\alpha$ & $q$ & $T_*$ & $\delta S_1$  &$\delta S_3$ &$T_*(\delta S_3-\delta S_1)$ & $\int^{\delta S_3}_{\delta S_1}Td\delta S$ & relative error \\   \hline
 0      & 0.12        &0.448178       & 0.0006548237    & 0.0038527782 &0.0014332529 & 0.0014330141 & $0.0167 \%$ \\
 0.25      & 0.12          &0.490898     & 0.0017571468      & 0.0067818433 & 0.0024666135 & 0.0024663505& $0.0107\%$ \\
 0.50     & 0.06       &0.65793667      & 0.0075207833     & 0.1020304981 & 0.0621814070 & 0.0622185738 & $0.0597\%$ \\
0.75      & 0.06          &1.070886315     & 0.1595664149     &11.1997305875 &11.8227607278 & 11.8253260328 & $0.0217\%$
       \end{tabular*}
       {\rule{\temptablewidth}{2pt}}
       \end{center}
       \label{tb6}
       \end{table}

 \begin{table}[!h]
\tabcolsep 0pt
\caption{Numerical check of equal area law in $T-\delta S$ graph for $n=3, \theta_0=0.3, l=1, b=1, k=1$}
\vspace*{-12pt}
\begin{center}
\def\temptablewidth{1\textwidth}
{\rule{\temptablewidth}{2pt}}
\begin{tabular*}{\temptablewidth}{@{\extracolsep{\fill}}c||c||c||c||c||c||c||c}
 $\alpha$ & $q$ &  $T_*$ & $\delta S_1$  &$\delta S_3$ &$T_*(\delta S_3-\delta S_1)$ & $\int^{\delta S_3}_{\delta S_1}Td\delta S$ & relative error \\   \hline
0      & 0.12          &0.277496     & 0.0394539472      & 0.1608172794 & 0.0336778392 & 0.0337657232& $0.2603\%$ \\
 0.25      & 0.12          &0.3002889      & 0.0711468332  &0.2719042091 & 0.0602852116 & 0.0603586328 & $0.1216\%$ \\
 0.50      & 0.06          &0.4078042      & 0.2149538454    & 2.3059365730 & 0.8527115384 & 0.8534806310 & $0.0901\%$ \\
 0.75      & 0.06          &0.65401625      & 6.7549216837       & 74.3949477659 & 44.2376762082 & 44.2373479097 & $0.0007\%$
       \end{tabular*}
       {\rule{\temptablewidth}{2pt}}
       \end{center}
       \label{tb7}
       \end{table}

       \begin{table}[!h]
\tabcolsep 0pt
\caption{Numerical check of equal area law in $T-\delta S$ graph for $n=4, \theta_0=0.3, l=1, b=1, k=1$}
\vspace*{-12pt}
\begin{center}
\def\temptablewidth{1\textwidth}
{\rule{\temptablewidth}{2pt}}
\begin{tabular*}{\temptablewidth}{@{\extracolsep{\fill}}c||c||c||c||c||c||c||c}
 $\alpha$ & $q$ & $T_*$ & $\delta S_1$  &$\delta S_3$ &$T_*(\delta S_3-\delta S_1)$ & $\int^{\delta S_3}_{\delta S_1}Td\delta S$ & relative error \\   \hline
 0      & 0.12        &0.448178      & 0.0046156698    &0.0177350761 & 0.0058798293 &0.0058821069 & $0.0387\%$ \\
 0.25      & 0.12          &0.490898     & 0.0080413189     & 0.0369024703 & 0.0141678815 & 0.0141621430 & $0.0405\%$ \\
 0.50     & 0.06       &0.65793667      & 0.0145412501  & 0.3493654246 & 0.2202931024 & 0.2207464643 & $0.2054\%$ \\
0.75      & 0.06          &1.070886315     & 0.5398143877    & 24.4969431344 & 25.6553613215 &25.6488620105 & $0.0253\%$
       \end{tabular*}
       {\rule{\temptablewidth}{2pt}}
       \end{center}
       \label{tb8}
       \end{table}

\section{Conclusions}
\label{sec:7}
 In this paper, we probe the critical phenomena of dilaton AdS black holes from a totally different perspective other than the $P-v$ criticality and the $q-U$ criticality discussed in the former literature~\cite{zhaoren}.

 On the one hand, we study the two point correlation function of dilaton AdS black holes. Considering the points $(t_0,
x_i)$ and $(t_0, x_j)$ on the AdS boundary, one can write the equal time two point correlation function in the large $\Delta$ limit and obtain the proper length by parameterizing the trajectory with $\theta$. Applying the Euler-Lagrange equation, one can derive the equation of motion for $r(\theta)$. Solving this equation constrained by the boundary condition $r(0)= r_0, r'(0)=0$, we obtain $r(\theta)$ via numerical methods. To avoid the divergence, we regularize the geodesic length by subtracting the geodesic length $L_0$ in pure AdS with the same boundary region.

On the other hand, we investigate the entanglement entropy of dilaton AdS black holes which can be expressed holographically in terms of the area of a minimal surface $\Sigma$ anchored on $\partial A$ with boundary condition $\partial \Sigma=\partial A$. We choose the region to be a spherical cap on the boundary delimited by $\theta\leq\theta_0$ and obtain the holographic entanglement entropy through parametrizing the minimal surface by $r(\theta)$, Utilizing the Euler-Lagrange equation again, we derive the equation for $r(\theta)$. Solving the equation constrained by the boundary condition numerically, we obtain $r(\theta)$. For similar consideration, we regularize the entanglement entropy by subtracting the entanglement entropy $S_0$ in pure AdS with the same boundary region.

For both the two point correlation function and the entanglement entropy, we consider $4\times2\times2=16$ cases due to different choices of parameters. Namely, $\alpha$ is chosen as $0, 0.25, 0.5, 0.75$ respectively to probe the effect of dilaton gravity on the phase structure while $n$ is chosen as $3,4$ respectively to study the impact of spacetime dimensionality. $\theta_0$ is chosen as $0.2, 0.3$ respectively to check the effect of different boundary region sizes. The behavior of $T-\delta L$ graphs is quite similar to that of $T-\delta S$ graphs. Firstly, the van der Waals like behavior can be clearly witnessed from all the $T-\delta L$ ($T-\delta S$) graphs for $q<q_c$. Secondly, there exist both the local maximum temperature and the local minimum temperature. Thirdly, not only $T_{max}$ and $T_{min}$ but also the corresponding $\delta L$ ($\delta S$) increase as the increasing of the parameter $\alpha$, showing the impact of dilaton gravity. This phenomena is more apparent for large $\alpha$. Fourthly, $T_{max}$ and $T_{min}$ of the cases $n=4$ is higher than that of the cases $n=3$. Lastly, the effect of boundary region size is quite apparent in the range of $\delta L$ ($\delta S$) axis.

Furthermore, we discuss the stability of dilaton AdS black holes by introducing the analogous specific heat for $T-\delta L$ graph and $T-\delta S$ graph accordingly. All these graphs can be divided into three branches. Namely, the stable large radius branch with positive specific heat, the unstable medium radius branch with negative specific heat and the stable small radius branch with positive specific heat. This phenomena is quite similar to that of $T-S$ graph and $P-v$ graph of dilaton AdS black holes in former research. To remove the unstable branch, we introduce a bar $T=T_*$ vertical to the temperature axis. $T_*$ should be interpreted physically as the first order phase transition temperature and can be determined utilizing the free energy analysis. It is shown that the first order phase transition temperature $T_*$ is affected by both $\alpha$ and $n$. When $\alpha$ increases, $T_*$ increases. When $n$ increases, $T_*$ increases too. We also examine the analogous equal area law for both the $T-\delta L$ graph and the $T-\delta S$ graph respectively. The relative errors for all the cases are small enough that we can safely conclude that the analogous equal area law  holds for $T-\delta L$ ($T-\delta S$) graph of dilaton AdS black holes.

\acknowledgments  The author would like to express his sincere gratitude to Dr. Jieci Wang, Dr. Xiongjun Fang, Dr. Zixu Zhao, Prof. Jia-Lin Zhang, Prof. Qiyuan Pan, Prof. Songbai Chen and Prof. Jiliang Jing for their warm hospitality during the Summer School and Workshop on Gravitation and Cosmology held in Hunan Normal University. The author also want to thank Dr. Xiao-Xiong Zeng for helpful discussions on numerics.

\end{document}